\documentclass[twocolumn]{aastex631}

\usepackage{amsmath,amssymb,color}

\defcitealias{glazebrook17}{G17}
\defcitealias{schreiber18_dustsed}{S18a}
\defcitealias{Schreiber18_mosfire}{S18b}
\defcitealias{schreiber18_jekyllhyde}{S18c}
\defcitealias{schreiber21}{S21}

\submitjournal{ApJ}

\shorttitle{Low star-formation activity and low gas content of quiescent galaxies at $z=$ 3.5--4.0}
\shortauthors{Suzuki et al.}

\begin{document}

\title{Low star-formation activity and low gas content of quiescent galaxies at $z=$ 3.5--4.0 constrained with ALMA}

\correspondingauthor{Tomoko L. Suzuki}
\email{tomoko.suzuki@ipmu.jp}

\author[0000-0002-3560-1346]{Tomoko L. Suzuki}
\affiliation{Kavli Institute for the Physics and Mathematics of the Universe (WPI),The University of Tokyo Institutes for Advanced Study, The University of Tokyo, Kashiwa, Chiba 277-8583, Japan}

\author[0000-0002-3254-9044]{Karl Glazebrook}
\affiliation{Centre for Astrophysics and Supercomputing, Swinburne University of Technology, Hawthorn, VIC 3122, Australia}

\author[0000-0003-0942-5198]{Corentin Schreiber}
\affiliation{Department of Physics, University of Oxford, Clarendon Laboratory, Parks Road, Oxford OX1 3PU, UK}

\author[0000-0002-2993-1576]{Tadayuki Kodama}
\affiliation{Astronomical Institute, Tohoku University, 6-3, Aramaki, Aoba, Sendai, Miyagi 980-8578, Japan}

\author[0000-0003-1362-9302]{Glenn G. Kacprzak}
\affiliation{Centre for Astrophysics and Supercomputing, Swinburne University of Technology, Hawthorn, VIC 3122, Australia}
\affiliation{ARC Centre of Excellence for All Sky Astrophysics in 3 Dimensions (ASTRO 3D), Australia}

\author{Roger Leiton}
\affiliation{Departamento de Astronom\'ia, Universidad de Concepci\'on, Casilla 160-C, Concepci\'on, Chile}

\author[0000-0003-2804-0648]{Themiya Nanayakkara}
\affiliation{Centre for Astrophysics and Supercomputing, Swinburne University of Technology, Hawthorn, VIC 3122, Australia}

\author[0000-0001-5851-6649]{Pascal A. Oesch}
\affiliation{Department of Astronomy, University of Geneva, Ch. Pegasi 51, 1290 Versoix, Switzerland}
\affiliation{Cosmic Dawn Center (DAWN), Niels Bohr Institute, University of Copenhagen Jagtvej 128, K\o benhavnN, DK-2200, Denmark}

\author[0000-0001-7503-8482]{Casey Papovich}
\affiliation{Department of Physics and Astronomy, Texas A\&M University, College Station, TX, 77843-4242 USA}
\affiliation{George P.\ and Cynthia Woods Mitchell Institute for Fundamental Physics and Astronomy, Texas A\&M University, College Station, TX, 77843-4242 USA}

\author[0000-0001-5185-9876]{Lee Spitler}
\affiliation{Department of Physics and Astronomy, Macquarie University, Sydney, NSW 2109, Australia}
\affiliation{Research Centre in Astronomy, Astrophysics \& Astrophotonics, Macquarie University, Sydney, NSW 2109, Australia}
\affiliation{Australian Astronomical Optics, 105 Delhi Rd, North Ryde, NSW 2113, Australia}

\author[0000-0001-5937-4590]{Caroline M. S. Straatman}
\affiliation{Sterrenkundig Observatorium, Universiteit Gent, Krijgslaan 281 S9, B-9000 Gent, Belgium}

\author[0000-0001-9208-2143]{Kim-Vy Tran}
\affiliation{School of Physics, University of New South Wales, Kensington, Australia}

\author[0000-0002-2504-2421]{Tao Wang}
\affiliation{School of Astronomy and Space Science, Nanjing University, 163 Xianlin Avenue, Nanjing 210023, People’s Republic of China}
\affiliation{Key Laboratory of Modern Astronomy and Astrophysics (Nanjing University), Ministry of Education, Nanjing 210023, People’s Republic of China}



\begin{abstract}
The discovery in deep near-infrared surveys of a population 
of massive quiescent galaxies at $z>3$ has given rise to the question of how they came to be quenched so early in the history of the Universe.
Measuring their molecular gas properties can distinguish between physical processes where they stop forming stars due to a lack of fuel versus those where  star-formation efficiency is reduced and the gas is retained.
We conducted Atacama Large Millimeter/sub-millimeter Array (ALMA) observations of four quiescent galaxies at $z=$ 3.5--4.0 found by the Fourstar Galaxy Evolution Survey (ZFOURGE) and a serendipitous optically dark galaxy at $z=3.71$. 
We aim to investigate the presence of dust-obscured star-formation and their gas content by observing the dust continuum emission at Band-7 and the atomic carbon [{\sc C i}]($^3P_1$--$^3P_0$) line at 492.16~GHz. 
Among the four quiescent galaxies, only one source is detected in the dust continuum at $\lambda _{\rm obs} = 870~{\rm \mu m}$. 
The sub-mm observations confirm their passive nature, 
and all of them are located more than four times below the main sequence of star-forming galaxies at $z=3.7$.
None of the targets are detected in [{\sc C i}], constraining their gas mass fractions to be $<$ 20\%.
These gas mass fractions are more than three times lower than the scaling relation for star-forming galaxies at $z=3.7$.  
These results support scenarios where massive galaxies at $z=$ 3.5--4.0 quench by consuming/expelling all the gas rather than by reducing the efficiency of the conversion of their gas into stars. 
\end{abstract}



\section{Introduction} \label{sec:intro}

Galaxies show a clear bimodality in physical properties, with distinct star-forming and quiescent populations. 
This is seen in observations such as  the color--magnitude diagram \citep[e.g.,][]{baldry04}, the rest-frame {\it UVJ} diagram \citep[e.g.,][]{williams09}, and the stellar mass--star formation rate (SFR) diagram \citep[e.g.,][]{renzini_peng_15}. 
Galaxies are considered to transform, via a process known as  ``quenching'',  from young blue star-forming galaxies to older and redder galaxies with little or no star-formation. 
Revealing how such quenching occurs and changes across cosmic time is crucial to understand the physical mechanisms behind galaxy formation and evolution in the Universe.

Although various physical mechanisms have been proposed, it remains unclear what is the main driver of galaxy quenching \citep[see][and references therein]{man_belli18}. 
One of the key observables to distinguish quenching processes is molecular gas, because gas mass fraction ($f_{\rm gas} = M_{\rm gas}/(M_{\rm *} + M_{\rm gas})$)s
and gas depletion timescale ($t_{\rm dep} = M_{\rm gas}/{\rm SFR}$) are expected to change depending on the quenching mechanisms. 
In massive galaxies it is thought that a principle process is  feedback by an active galactic nucleus (AGN)
expelling  gas and thus stopping star-formation \citep[e.g.,][]{dimatteo05}. 
Galaxies could also reduce the star-formation activity by virial shock heating \citep[e.g.,][]{keres05,dekel09.nature} and halo gas stripping (i.e., starvation mechanisms; \citealt{larson1980,peng15}). 
In both cases, the cold gas cannot fall into galaxies anymore and galaxies quench when they consume all their remaining gas.  
Gas-rich major mergers or violent disk instabilities can induce starbursts with short gas depletion timescales which quench galaxies by quickly consuming all the gas \citep[e.g.,][]{dekelburkert14}. 
Morphological transformation, such as the formation of a central bulge, can dynamically stabilise gas against star formation \citep{martig09}  which would result in quiescent galaxies with high gas fractions and long gas depletion times.

Thus observations of molecular gas and star formation in quiescent galaxies can narrow the range of possibilities of quenching mechanisms. 
Studies of the gas properties of quiescent galaxies have been conducted up to $z\sim3$ 
directly using CO lines \citep{sargent15,spilker18,belli21,caliendo21,williams21}
or indirectly using dust continuum emission \citep{gobat18, magdis21,whitaker21}. 
Some of these studies revealed low gas mass fractions ($f_{\rm gas}\lesssim0.1$) and relatively short gas depletion timescales ($t_{\rm dep} \lesssim 1$~Gyr) of quiescent galaxies at $z\sim$ 0.5--3.0, suggesting that these galaxies quench by strong feedback or rapid gas consumption due to active star-formation \citep[e.g.,][]{sargent15,spilker18,whitaker21}.
\citet{gobat18} obtained a gas mass fraction of $\sim0.1$ and a longer gas depletion timescale of 2--3~Gyr for quiescent galaxies at $z\sim1.8$ from a stacking analysis, suggesting that the quenching is caused by the reduced star-formation efficiency.
The reported gas properties of quiescent galaxies at $z\gtrsim1$ appear to show a large variety  \citep{spilker18,belli21,williams21}, and there is no consensus on the typical gas properties of quiescent galaxies at high redshift yet.

In the last few years massive quiescent galaxies at $z>3$ have been confirmed using near-infrared (NIR) spectroscopic observations \citep{glazebrook17,Schreiber18_mosfire,tanaka19,forrest20,valentino20,kubo21}. 
Investigating the gas properties of quiescent galaxies at $z>3$, where there is less cosmic time for quenching to happen, allows insight into the range of physical quenching mechanisms as the quenching mechanisms operate on different timescales.
Observations at sub-millimeter (mm) wavelengths are necessary to investigate the molecular gas contents of quiescent galaxies at $z>3$.
Furthermore, because the current level of star-formation in these galaxies is primarily constrained by rest-frame optical emission lines and spectral energy distribution (SED) fitting with the optical--NIR data, observations at longer wavelengths are needed to directly confirm whether they have dust obscured star-formation activity \citep{simpson17,schreiber18_jekyllhyde,santini19}.

CO molecular lines have been used to measure the gas masses of quiescent galaxies at high redshifts  \citep{sargent15,spilker18,belli21,caliendo21,williams21}. 
At $z>3$, mid/high-{\it J} CO lines ($J_{\rm upper}\ge4$) are only observable with Atacama Large Millimeter/sub-millimeter Array (ALMA). 
One of the major uncertainties when calculating molecular gas masses from mid/high-{\it J} CO lines is the CO gas excitation state, 
which has been investigated only for dust-obscured galaxies such as sub-mm galaxies (SMGs) especially at $z>3$ \citep[e.g.,][]{riechers20,liu21}.
Atomic carbon [{\sc C i}]($^3P_1$--$^3P_0$) line ($\nu_{\rm rest}=492.16$~GHz) has been proposed as a good tracer of molecular gas in galaxies with advantages over CO molecular lines at higher redshift \citep[e.g.,][]{papadopoulos_thi_viti04,bisbas17}. 
Previous observational studies have reported  [{\sc C i}] detections from normal star-forming galaxies as well as SMGs at $z>1$  \citep[e.g.,][]{walter11,bothwell17,popping17,valentino20_CI}.
Given that no assumption on the CO gas excitation state is required, the [{\sc C i}] line can be a good tracer of molecular gas for quiescent galaxies at high redshift and has the additional advantage that it can reduce the uncertainty on the molecular gas mass measurement.

In this paper, we report results obtained from ALMA observations of quiescent galaxies at $z=$ 3.5--4.0, and one optically dark galaxy at $z=3.71$ which is a companion of one of the quiescent galaxies and is in a transition phase to quiescence \citep[][hereafter, S18c and S21]{schreiber18_jekyllhyde,schreiber21}. 
We investigate their dust-obscured star-formation activity and molecular gas content by observing the dust continuum emission at $\lambda_{\rm obs}=870\mu{\rm m}$ and the [{\sc C i}] line, respectively. 
We aim to confirm the passive nature of the quiescent galaxies at sub-mm wavelengths and to understand how massive galaxies at $z>3.5$ stop star-formation by deriving a constraint on their molecular gas properties.

This paper is organized as follows. 
In Section~\ref{sec:zfourge}, we briefly explain the target galaxies. 
In Section~\ref{sec:obs}, we describe the ALMA observations and explain the data reduction and stacking analyses. 
We show our results on the star-formation activity and gas properties of the quiescent galaxies at $z=$ 3.5--4.0 in Section~\ref{sec:result}. 
We summarize the main findings in this study in Section~\ref{sec:summary}. 
Throughout this paper, we use the cosmological parameters: 
$H_0 = 70\ {\rm km\ s^{-1}\ Mpc^{-1}}$, $\Omega_{\rm m} = 0.3$ and $\Omega_\Lambda = 0.7$. 
We assume the \citet{chabrier03} initial mass function (IMF).

\section{Targets}\label{sec:zfourge}

\subsection{Quiescent galaxies at $z>3.5$ from ZFOURGE}\label{subsec:zfourge_qg}

Our main targets are four spectroscopically confirmed quiescent galaxies at $z=$ 3.5--4.0 selected from the Fourstar Galaxy Evolution Survey (ZFOURGE; \citealt{straatman16}), namely, ZF-COSMOS-20115 (also called `Jekyll' in \citetalias{schreiber18_jekyllhyde}), ZF-UDS-8197, ZF-COS-18842, and ZF-COS-19589.  
They were classified as quiescent galaxies based on the rest-frame {\it UVJ} colors and subsequently observed with Keck/MOSFIRE \citep{glazebrook17,Schreiber18_mosfire}. 
In this section, we briefly summarize the results from \citet[][hereafter, G17]{glazebrook17} and \citet[hereafter, S18b]{Schreiber18_mosfire} for the four quiescent galaxies. 
We refer the reader to these studies for more details.

Jekyll is a massive quiescent galaxy at $z_{\rm spec} = 3.715$ with strong Balmer absorption lines, 
which was initially spectroscopically confirmed in  \citetalias{glazebrook17}. 
ZF-UDS-8197 is spectroscopically confirmed as $z_{\rm spec}$ = 3.547 with the [{\sc Oiii}]$\lambda\lambda$5007,4959 doublet (\citetalias{Schreiber18_mosfire}). 
The [{\sc Oiii}] emission line in the MOSFIRE spectrum of this source has a broad line width of $\sigma_{v} = 530\pm53\ {\rm km\ s^{-1}}$. 
\citetalias{Schreiber18_mosfire} argued that such a broad [{\sc Oiii}] emission line is likely to be emitted from shock-excited gas rather than from narrow-line regions of the AGN.
ZF-COS-18842 and ZF-COS-19589 are classified as sources with uncertain $z_{\rm spec}$ in \citetalias{Schreiber18_mosfire}. 
ZF-COS-18842 is identified with a single emission line with $5.6\sigma$, 
which is attributed to the [{\sc Oiii}]$\lambda$5007 line.  
ZF-COS-19589 is assigned $z_{\rm spec}=3.715$ with a low probability of $p=32\%$ based on weak Balmer absorption features. 
Given that this redshift lies within $\Delta z < 0.01$ of Jekyll and this source is located only 23~arcsec away from Jekyll on the sky,  there is a strong possibility that this galaxy is  physically associated with Jekyll. 
Note that the spectroscopic redshifts of ZF-COS-18842 and ZF-COS-19589 are consistent with their photometric redshifts \citepalias{Schreiber18_mosfire}. 
None of these four galaxies is detected with X-ray observations either by {\it Chandra} \citep{civano16,marchesi16} or {\it XMM-Newton} (XMM-COSMOS: \citealt{cappelluti07,hasinger07,brusa10}, Subaru-XMM Newton Deep Survey: \citealt{akiyama15}). 
\citetalias{glazebrook17} and \citetalias{Schreiber18_mosfire} conducted SED fitting analyses for the quiescent galaxies with multi-wavelength photometry and MOSFIRE spectra to obtain their global physical quantities and star-formation histories. 
Their basic global physical quantities obtained by \citetalias{Schreiber18_mosfire} are summarized in Table~\ref{tab:obssummary}. 
The star-formation histories inferred from the SED fitting indicate that the four quiescent galaxies have formed at $z\sim$ 5.1--6.7, 
experienced a main star-formation period for 0.1--0.7~Gyr and then quenched 0.3--0.5~Gyr prior to the epoch of observation \citepalias{Schreiber18_mosfire}.

\subsection{Hyde: an optically dark galaxy at $z=3.71$}\label{subsec:target_hyde}
Jekyll is accompanied by a source with sub-mm emission at a distance of $\sim0''.4$ \citep{simpson17}. 
Follow-up observations with ALMA Band-8 (385--500~GHz) for Jekyll found that this sub-mm emission comes from another extremely dust-obscured galaxy at $z_{[\textsc{Cii}]}$=3.709 \citepalias{schreiber18_jekyllhyde}.   
This optically-dark galaxy at $z=3.709$  was dubbed `Hyde'. 
Detailed studies for Hyde with ALMA were performed by \citetalias{schreiber18_jekyllhyde} and \citetalias{schreiber21}. 
The physical quantities of Hyde are characterized as follows: 
$\rm log(M_*/M_\odot)=10.90_{-0.37}^{+0.21}$, 
$T_{\rm dust}=31\pm3$~K, and ${\rm SFR_{IR}}=50_{-18}^{+24}\, {\rm M_\odot\, yr^{-1}}$. 

Hyde is considered to be a  galaxy transitioning to quiescence because of its low SFR as compared to typical main sequence star-forming galaxies with similar stellar masses at $z\sim3.7$ (\citetalias{schreiber21} and Section~\ref{subsec:SFRIR}). 
Because Hyde could be representative of an important population relevant to understanding the quenching of massive galaxies at $z>3.5$, we investigate its gas properties together with the four quiescent galaxies in this study.

\begin{table*}[]
\caption{Summary of the observed and physical quantities of the four quiescent galaxies at $z=$ 3.5--4.0 and Hyde \citepalias{schreiber21}. All the galaxies are observed with ALMA Band-7 and Band-3.}
\begin{center}
\begin{tabular}{llcccccccccc} \hline
ID & $z_{\rm spec}$ & $\rm log(M_*/M_\odot)$  &  $S_{\rm 870\mu {\rm m}}$ & $S_{\rm [CI]} \Delta v ^{a}$ & \multicolumn{2}{c}{${\rm log(L_{IR}/L_\odot)}^b$} & \multicolumn{2}{c}{${\rm SFR_{IR}}^b$} &  $\rm log(M_{\rm gas}/M_\odot)$  & $f_{\rm gas}$ \\
 & &  & [mJy] & [Jy $\rm km\ s^{-1}$] & \multicolumn{2}{c}{} & \multicolumn{2}{c}{[$\rm M_\odot\ yr^{-1}$]} & \\ \hline 
 & &  & & & 40~K & 20~K & 40~K & 20~K & \\ \hline
ZF-COS-19589 & 3.715$^c$ & $10.79_{-0.06}^{+0.07}$ & 0.21$\pm$0.04  & $<$0.06 & $11.51_{-0.09}^{+0.08}$ & $10.74_{-0.09}^{+0.08}$ & $33_{-7}^{+6}$ & 6$\pm$1 & $<$10.15 & $<$0.19 \\
ZF-COS-18842 & 3.782$^{c}$ & $10.65_{-0.04}^{+0.06}$  &  $<$0.13        &  $<$0.05 & $<$11.29 & $<$10.53  & $<$20 & $<$3 & $<$10.05 & $<$0.20 \\
ZF-UDS-8197 & 3.543        & $10.56_{-0.05}^{+0.05}$  &  $<$0.10        & $<$0.04  & $<$11.18 & $<$10.39 & $<$15 & $<$2 & $<$9.90 & $<$0.18 \\
Jekyll & 3.715 & $11.06_{-0.03}^{+0.06}$        &  $<$$0.05^{d}$          & $<$0.11 & $<$11.06 & $<$10.21 & $<$12 & $<$2 & $<$10.38 & $<$0.17 \\ \hline 
Hyde   & 3.709    &     $10.90_{-0.37}^{+0.21}$   &  $1.10$$\pm$$0.02^{d}$  & $<$0.07 & \multicolumn{2}{c}{$12.03_{-0.14}^{+0.14}$$^e$} & \multicolumn{2}{c}{$50_{-18}^{+24}$$^e$} & $<$10.20 & $<$0.17 \\ \hline 
\end{tabular}
\end{center}
\tablecomments{
All of the upper limits are $3\sigma$ upper limits. 
$^a$We assumed a velocity width of $400\, {\rm km\, s^{-1}}$ except for Hyde. 
For Hyde, a velocity width of $800\, {\rm km\, s^{-1}}$ was assumed \citepalias{schreiber18_jekyllhyde}. 
$^b$We show both values assuming $T_{\rm dust}=40$~K and 20~K except for Hyde. 
$^c$Uncertain redshifts \citepalias{Schreiber18_mosfire}. 
$^d$The central wavelength is $992\mu{\rm m}$.  
$^e$Values obtained by \citetalias{schreiber21}. The IR luminosity corresponds to the bolometric IR luminosity including the contribution from the dust heating by old stellar components. 
}
\label{tab:obssummary}
\end{table*}%

\section{Observations and analyses}\label{sec:obs}

All the targets including Hyde were observed with ALMA Band-7 (275--373~GHz) and Band-3 (84--116~GHz) 
in observing programs in Cycle~6 (2018.1.00216.S; PI: C. Schreiber) and 7 (2019.1.01329.S; PI: T. Suzuki).

\subsection{Band-7 observation to trace dust continuum}\label{subsec:b7data}

The Band-7 observations of three quiescent galaxies, namely, ZF-COS-19589, ZF-COS-18842, and ZF-UDS-8197, were conducted in October 2019 as a part of the Cycle~7 observing program.
The integration time is 16--17~min for the three targets.
The central frequency of four spectral windows was set to be $343.5$~GHz ($\sim870{\rm \mu m}$). 
We calibrated the raw data using the Common Astronomy Software Application package ({\sc casa}; \citealt{CASA}). 
We ran the {\sc clean} algorithm with Briggs weighting (robust parameter$=$0.5). 
When there are sources with $\ge5\sigma$ significance in the constructed maps, we ran {\sc clean} again by masking the sources. 
The final sensitivity of the continuum maps at $870\mu{\rm m}$ 
is $0.04$~mJy/beam for ZF-COS-19589 and ZF-COS-18842 and $0.03$~mJy/beam for ZF-UDS-8197. 
The average beam size is $0''.40\times0''.35$.

The Band-7 data for Jekyll and Hyde was taken as a part of the Cycle~6 program. 
Four spectral windows were set so that the [{\sc Nii}] emission line at $\nu_{\rm rest}=1461.13$~GHz from Hyde can be covered \citepalias{schreiber21}. 
The integration time is 48~min.
We created the Band-7 continuum map for Jekyll and Hyde using the three spectral windows free from the emission line.
The observed frequency is $302.34$~GHz ($991.6\mu{\rm m}$). 
The achieved sensitivity is 0.02~mJy/beam and the beam size is $0''.24\times0''.20$.

\begin{figure}
    \centering\includegraphics[width=1.0\columnwidth]{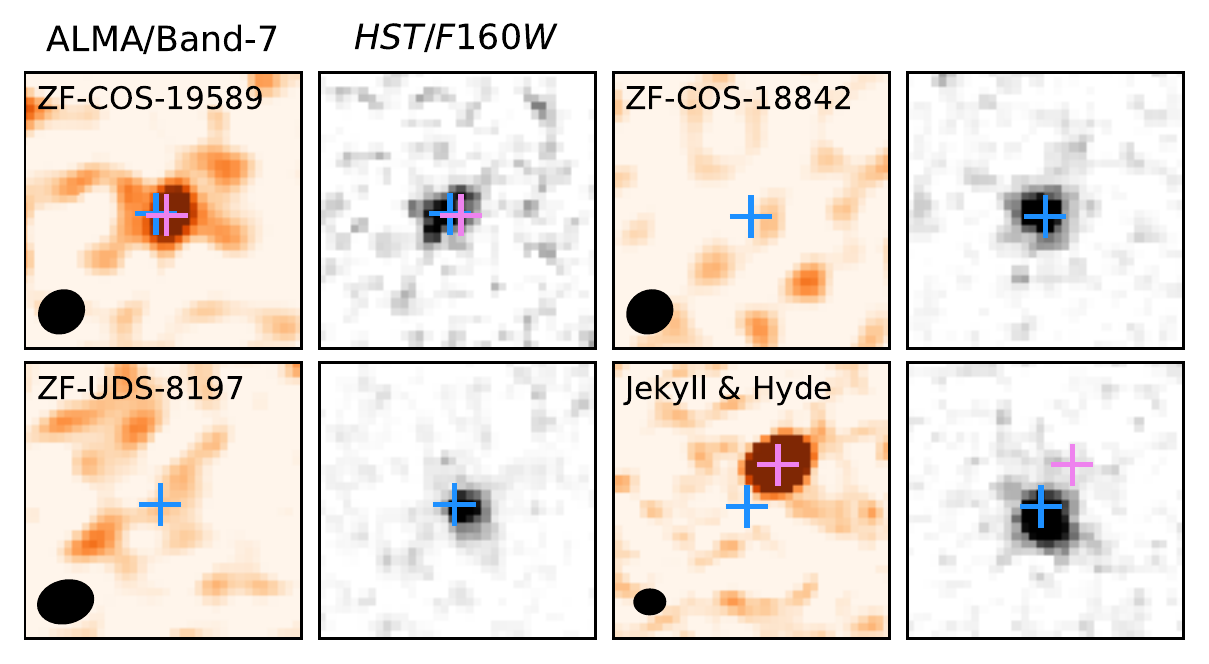}
    \caption{ALMA Band-7 and {\it HST}/{\it F160W} images of all the targets (image size: $3''\times3''$).  
    The blue and magenta plus marks show the coordinates from the ZFOURGE catalog \citep{straatman16} and the position of the $870\mu{\rm m}$ flux peak (when detected), respectively. 
    The black filled circle in each ALMA map represents the beam size.}
    \label{fig:thumbnail}
\end{figure}

The Band-7 continuum maps of all the targets are shown in Figure~\ref{fig:thumbnail} together with the {\it Hubble Space Telescope} WFC3/{\it F160W} images from CANDELS \citep{grogin11}. 
We conducted the source detection and flux measurement for the continuum maps using the {\sc casa} task {\sc imfit}. 
We fixed the central positions to the coordinates from the ZFOURGE catalog, which were determined with optical--NIR images \citep{straatman16}. 
Our detection criterion is that the peak flux from {\sc imfit} has $\ge3\sigma$ significance as done in \citet{suzuki21}. 
ZF-COS-19589 and Hyde are detected with $>5\sigma$. 
The peak flux obtained with {\sc imfit} is regarded as a total dust continuum flux in the following analyses. 
The other three quiescent galaxies are not detected with $>3\sigma$, and we assigned $3\sigma$ upper limits. 
The obtained continuum fluxes and upper limits are summarized in Table~\ref{tab:obssummary}.

The offset between the coordinate from the ZFOURGE catalog and that of the $870\mu{\rm m}$ continuum flux peak for ZF-COS-19589 is $0''.08$.
This is similar as a typical positional offset between ALMA sources and their {\it Ks}-band counterparts found in the COSMOS field reported in  \citetalias{schreiber18_jekyllhyde}. 
The observed positional offsets are considered to come from the positional accuracy of ALMA and VISTA/{\it Ks}-band images determined by the S/N of a source and the size of beam or point spread function \citepalias{schreiber18_jekyllhyde}.
We can clearly infer that the detected continuum emission is associated to ZF-COS-19589, unlike the case of Jekyll and Hyde (Figure~\ref{fig:thumbnail}) as discussed in \citetalias{schreiber18_jekyllhyde}.

\subsection{Band-3 observation to detect the [CI] line}\label{sec:b3data}

In the Cycle~7 project, we observed the four quiescent galaxies with Band-3 to detect the [{\sc C i}]($\rm ^3P_1-^3P_0$) line ($\nu_{\rm rest}=492.16$~GHz). 
The Band-3 observations of Jekyll and ZF-COS-19589 were conducted in October 2019. 
The observations of ZF-COS-18842 were conducted in March 2020 and March 2021. 
ZF-UDS-8197 was observed in March 2020, April, and June 2021. 
We set the frequencies of the spectral windows so that we  cover the [{\sc C i}] line from the targets. 
The integration time is 29, 80, 187, and 234~min for Jekyll, ZF-COS-19589, ZF-COS-18842, and ZF-UDS-8197, respectively.
The minimum spectral resolution is set to be 7.81~MHz.
We calibrated the raw data using {\sc casa} and created the data cubes. 
The RMS level over $400\, {\rm km\,s^{-1}}$ velocity width of the data cube is 0.053, 0.041, 0.032, and 0.091 $\rm mJy\, beam^{-1}$ for ZF-COS-19589, ZF-COS-18842, ZF-UDS-8197, and Jekyll, respectively.
The data cubes of ZF-COS-19589, ZF-UDS-8197, and Jekyll have similar beam sizes of $\sim 0''.60 \times 0''.50$. 
The beam size of ZF-COS-18842 is $0''.95\times0''.80$. 

The Band-3 observations of Hyde were conducted in September 2019 as part of the Cycle~6 program. 
The integration time is 198~min, 
and the minimum spectral resolution is set to be 7.81~MHz.
The spectral data cube was created with {\sc casa}/{\sc tclean} after subtracting the continuum emission. 
The RMS level measured over $800\ {\rm km\ s^{-1}}$ width is 0.030~$\rm mJy\ beam^{-1}$. 
We use the velocity width of $800\ {\rm km\ s^{-1}}$ for Hyde, which comes from the [{\sc Cii}] line width measured by \citetalias{schreiber18_jekyllhyde}. 
The beam size of the data cube is $0''.42\times0''.27$.

\begin{figure}
    \centering\includegraphics[width=1.0\columnwidth]{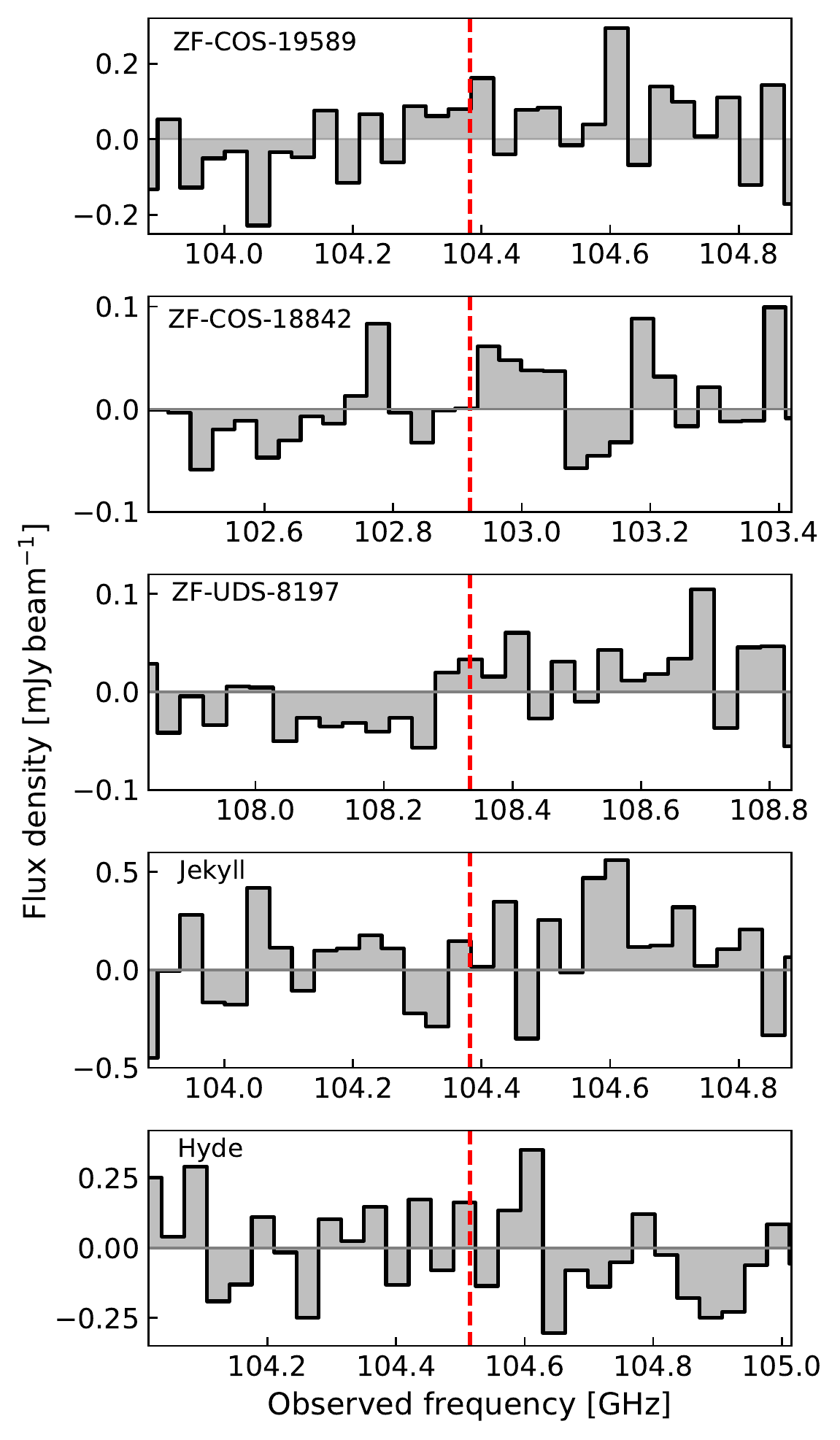}
    \caption{1D spectrum around the [{\sc C i}] line extracted at the position from the ZFOURGE catalog of each target. The velocity binning size is $\rm 100\, km\, s^{-1}$. The red dashed line in each panel corresponds to the frequency of the redshifted [{\sc C i}] line. 
    We found no emission line feature with $>3\sigma$ at the observed frequency of [{\sc C i}].}
    \label{fig:CIspectra}
\end{figure}

For each target, we created the data cubes with velocity widths of 100, 200, 400, and 800\, ${\rm km\,s^{-1}}$.  
We then ran the {\sc imfit} task around the frequency corresponding to the redshifted [{\sc C i}] line by fixing the central positions as done for the Band-7 data. 
We found no line feature with $>3\sigma$ in the four data cubes with different velocity widths for any of the sources.
Figure~\ref{fig:CIspectra} shows the one-dimentional (1D) spectra extracted at the optical--NIR positions.
We assign the $3\sigma$ upper limits on the [{\sc C i}] line flux based on the RMS measured over $400\, {\rm km\,s^{-1}}$ velocity width for the four quiescent galaxies and $800\, {\rm km\,s^{-1}}$ velocity width for Hyde. 
The $3\sigma$ upper limits on the velocity-integrated [{\sc C i}] line fluxes are summarized in Table~\ref{tab:obssummary}. 

\subsection{Stacking of the dust continuum maps}\label{subsec:stacking_cont}
We conducted a stacking analysis for the Band-7 continuum maps of the two non-detected quiescent galaxies, namely, ZF-COS-18842 and ZF-UDS-8197, with similar stellar masses and RMS levels.
We stacked the two continuum maps by weighting each map according to its RMS level. 
The central position of each map is fixed at the coordinate from the ZFOURGE catalog. 
The RMS level of the stacked continuum map is $0.026$~mJy.
We conducted the source detection with {\sc imfit} at the center of the stacked map as done for the individual galaxies (Section~\ref{subsec:b7data}), and found no emission with $\ge3\sigma$. 
The dust continuum flux upper limit for the stacked sample is $<0.078$~mJy ($3\sigma$).

\subsection{Stacking of the [CI] data cubes}\label{subsec:stacking_ci}
We conducted a stacking analysis for the [{\sc C i}] line of the five galaxies including Hyde. 
First of all, we created the {\it uv}-tapered maps of the sources except for ZF-COS-18842 to create the data cubes with a beam size of $\sim 1''.0 \times 0''.9$. 
We consider that all the five sources are at the $z_{\rm spec}$ determined by \citetalias{Schreiber18_mosfire}. 
We then stacked the data cubes of the five sources. 
We here used the data cubes with the velocity width of 100, 200, and 400 $\rm km\, s^{-1}$ \citep{spilker18}. 
When stacking, the individual spectra were shifted to the rest-frame frequency and weighted according to the RMS level of the (tapered) data cubes at the frequency of the [{\sc C i}] line. 

The source detection in the stacked data cubes was done as described in Section~\ref{sec:b3data}. 
There is no detection with $\ge3\sigma$ in the three stacked data cubes with different velocity widths. 
We determined the upper limit on the [{\sc C i}] line luminosity using the data cube with $\rm 400\, km\, s^{-1}$ width. 
The [{\sc C i}] line luminosity upper limit for the stacked sample is ${\rm log}(L_{\rm [CI](1-0)}/L_\odot) < 6.41$ ($3\sigma$). 

Note that only Jekyll has the stellar mass of $\rm log(M_*/M_\odot) >11$ among our targets (Table~\ref{tab:obssummary}). 
Given the stellar mass dependence of gas mass fraction \citep[e.g.,][]{tacconi18}, 
the stacked [{\sc C i}] line luminosity obtained above, and thus, the gas mass estimated with the stacked [{\sc C i}] line luminosity (Section~\ref{subsec:result_fgas}) may be inappropriate for the less massive galaxies. 
In order to check this, we stacked the Band-3 data cubes for the four galaxies except for Jekyll. 
We confirm that the [{\sc C i}] line luminosity upper limit, and thus, the gas mass upper limit for the stacked sample (Section~\ref{subsec:result_fgas}) do not largely change depending on whether Jekyll is included or not. 
This means that the stacking result for the five galaxies is not heavily weighted toward Jekyll and appropriate for the less massive galaxies.

\section{Results and discussion}\label{sec:result}

\begin{figure*}[t]
    \centering\includegraphics[width=1.0\textwidth]{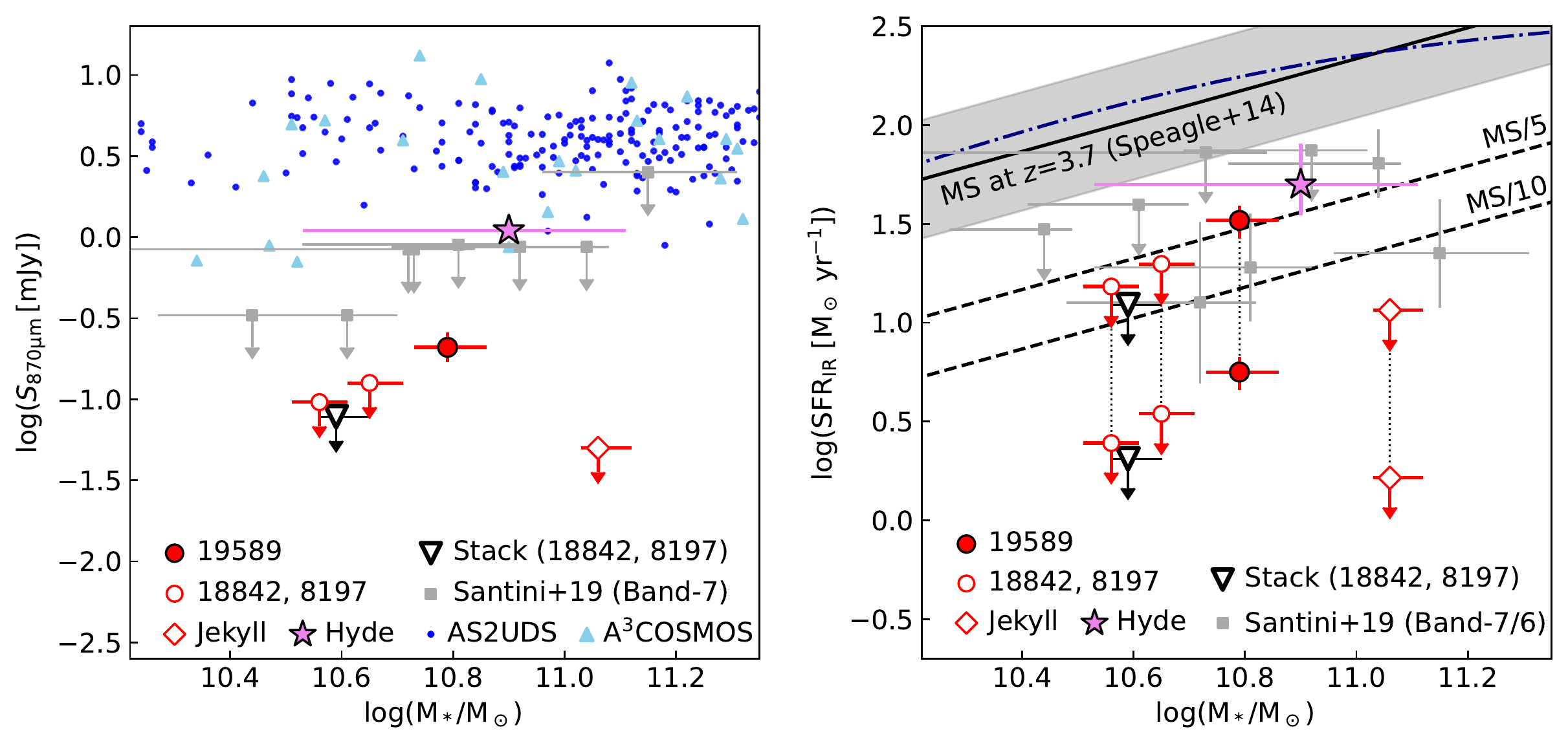}
    \caption{(Left) Continuum flux at $\sim870\, \mu{\rm m}$ as a function of stellar mass of our targets. For comparison, we show 
    galaxies with the individual Band-7 detections at $z=$ 3--4 from the literature (AS2UDS; \citealt{dudzeviciute20}, $\rm A^3COSMOS$; \citealt{liu19_I}) and 
    quiescent galaxies at $z=$ 3--5 with Band-7 data \citep{santini19}. 
    The four quiescent galaxies in this work are fainter in sub-mm by $\gtrsim1$~dex than SMGs at $z=$ 3--4. 
    (Right) $\rm SFR_{IR}$ as a function of stellar mass of our targets together with quiescent galaxies at $z=$ 3--5 with Band-7 and/or 6 data from \citet{santini19}.
    We show the SFRs estimated with both $T_{\rm dust}=$ 40~K and 20~K for each source except for Hyde. 
    The data points with the lower SFRs represent the results assuming $T_{\rm dust}=$ 20~K. 
    The black solid line shows the main sequence of star-forming galaxies at $z=3.7$ from \citet{speagle14}. 
    The dash-dotted line shows the star-forming main sequence at $z=3.7$ from \citet{tomczak16}, which is established with the ZFOURGE galaxy catalog. 
    The four quiescent galaxies including ZF-COS-19589 with the continuum detection have more than four times lower SFRs than the main sequence galaxies even when assuming the higher $T_{\rm dust}$. All the upper limits are $3\sigma$ upper limits. 
    }
    \label{fig:massSFR}
\end{figure*}

\subsection{$870\mu{\rm m}$ flux density}\label{subsec:result_cont}

In the left panel of Figure~\ref{fig:massSFR}, we compare the $\sim$870~$\mu{\rm m}$ continuum fluxes of our targets at $z\sim3.7$ with galaxies at $z=$ 3--4 individually detected with Band-7 from the literature \citep{liu19_I,dudzeviciute20} and quiescent galaxies at $z=$ 3--5 with Band-7 data from \citet{santini19}.
\citet{santini19} investigated the dust-obscured star-formation activity of their photometrically identified quiescent galaxies with ALMA archival data (Band-6 and 7). 
In Figure~\ref{fig:massSFR}, we show those quiescent galaxies that were confirmed with high confidence \citep{santini19}.

We find that our four quiescent galaxies, irrespective of being detected with Band-7 or not, have $\sim1$~dex fainter $\sim870\mu{\rm m}$ continuum fluxes than the SMGs at $z=$ 3--4. 
Hyde also has a fainter sub-mm flux than most of the SMGs with similar stellar masses. 
This indicates low dust-obscured star-formation activities in our targets as discussed in more details in the next section.  
By conducting a deeper observation for a spectroscopically confirmed sample than the previous work \citep{santini19}, we can give strong upper limits on the dust continuum emission and thus $\rm SFR_{IR}$ of quiescent galaxies at $z>3.5$.

\subsection{Dust-obscured star-formation activity}\label{subsec:SFRIR}
  
We converted the $\sim 870\mu{\rm m}$ continuum flux to a total IR luminosity ($L_{\rm IR}$) using a dust SED library by \citet[][hereafter, S18a]{schreiber18_dustsed}. 
The required parameters are the dust temperature ($T_{\rm dust}$) and the mid-to-total IR color, ${\rm IR8} \equiv L_{\rm IR}/L_{8 {\rm \mu m}}$. 
The SED templates have a dust emissivity index of $\beta\simeq1.5$ \citepalias{schreiber18_dustsed}. 
We assume $\rm IR8=7.37$, a typical value of star-forming galaxies at $z>2$ \citepalias{schreiber18_dustsed}. 
When we change the IR8 value by $-0.5\, (+0.5)$~dex, the change of $L_{\rm IR}$ is only $+0.06\, (-0.02)$~dex. 
This means that this parameter has a much smaller impact on the $L_{\rm IR}$ estimates than $T_{\rm dust}$ as discussed below.

Dust temperature is one of the major uncertainties when estimating $L_{\rm IR}$. 
Because it is difficult to estimate $T_{\rm dust}$ of our targets with the available data, we here assume two dust temperatures, namely, $T_{\rm dust}=40$~K and $20$~K.  
$T_{\rm dust}=40$~K is a typical value of star-forming galaxies at $z=$ 3.5--4.0 from the relation between redshift and $T_{\rm dust}$ given by \citetalias{schreiber18_dustsed}. 
$T_{\rm dust} = 20$~K is motivated by recent observations of quiescent galaxies up to $z\sim2$ \citep{gobat18,magdis21}. 
These studies reported that quiescent galaxies tend to have lower dust temperatures than star-forming galaxies at the same epoch.

By comparing the observed flux densities at Band-7 and the model SED normalized with $L_{\rm IR}$ from \citetalias{schreiber18_dustsed}, we calculated $L_{\rm IR}$ of the four quiescent galaxies. 
Using the two dust temperatures, we obtained two different estimates of $L_{\rm IR}$ (both are included in Table~\ref{tab:obssummary}). 
The $3\sigma$ upper limits on $L_{\rm IR}$ obtained from the stacking analysis (Section~\ref{subsec:stacking_cont}) are ${\rm log}(L_{\rm IR}/L_\odot)<$ 11.09 and 10.31 with $T_{\rm dust}=$ 40~K and 20~K, respectively.

As a test, we also calculated $L_{\rm IR}$ with a IR SED template ($T_{\rm dust}\sim20$~K) obtained by stacking analyses for massive quiescent galaxies at $z<2$ in \citet{magdis21}. 
The obtained $L_{\rm IR}$ are broadly consistent with those obtained with the IR SED template from \citetalias{schreiber18_dustsed} assuming $T_{\rm dust}=$ 20~K. 
The difference between the two measurements is less than 0.1~dex. 
This means that the applied IR SED templates do not significantly affect our results on $L_{\rm IR}$, and thus $\rm SFR_{IR}$, under a similar {$T_{\rm dust}$}.

We converted $L_{\rm IR}$ to SFR using the equation from \citet{kennicutt98}. 
We divided the SFRs by a factor of 1.7 to take into account the difference between the \citet{salpeter55} IMF and \citet{chabrier03} IMF \citep{pozzetti07}. 
The right panel of Figure~\ref{fig:massSFR} shows the relation between stellar mass ($M_{*}$) and $\rm SFR_{IR}$ for the four quiescent galaxies and Hyde. 
The $\rm SFR_{IR}$ of Hyde was estimated in \citetalias{schreiber21} using the same IR SED library. 
The $3\sigma$ upper limit on $\rm SFR_{IR}$ of the stacked sample becomes $< 12.3\, \rm M_\odot\, yr^{-1}$ and $< 2.1\, \rm M_\odot\, yr^{-1}$ with $T_{\rm dust}=$ 40~K and 20~K, respectively.

In the right panel of Figure~\ref{fig:massSFR}, we find that all the galaxies and the stacking result are located below the $M_*$--SFR relation of star-forming galaxies (the so-called star-forming main sequence) at $z=3.7$ \citep{speagle14,tomczak16} by a factor of $>4$. 
This result is valid irrespective of the assumed $T_{\rm dust}$. 
With $T_{\rm dust}=20$~K, the SFRs of the four quiescent galaxies are more than 20 times lower than those of star-forming galaxies on the main sequence at $z\sim3.7$. 
Interestingly, ZF-COS-19589, which is detected at $870\, {\rm \mu m}$, still appears to have a significantly lower SFR than the main sequence galaxies at the same redshift. 
Given that ZF-COS-19589 is associated to dust emission but has a weak star-formation activity like Hyde, ZF-COS-19589 may  be a similar transitioning galaxy  \citepalias{schreiber21}.

\citetalias{Schreiber18_mosfire} measured the SFRs of the four quiescent galaxies with several methods, namely, the optical--NIR SED fitting, H$\beta$ and [{\sc Oii}] emission line flux estimated from the spectral fitting. 
The $\rm SFR_{IR}$ (upper limits) obtained in this study are consistent with the SFRs estimated with the other methods within the uncertainties. 
Our ALMA observations confirm the low star-formation activity of the four quiescent galaxies at the sub-mm wavelengths in addition to at the optical--NIR wavelengths \citepalias{Schreiber18_mosfire}.

We note that the old stellar populations in quiescent galaxies can be an additional heating source of dust in the general interstellar medium (ISM). 
The old stellar populations in ISM can produce IR emission even in the absence of young and massive stars heating dust in birth clouds. 
Including the IR emission heated by old stars can lead to an overestimate of $\rm SFR_{IR}$  \citep[e.g.,][]{dacunha08}. 
Indeed, \citetalias{schreiber21} estimated that in Hyde, 60\% of the total $L_{\rm IR}$ is contributed to by old stars. 
However, the contributions from old stars, if any, do not significantly change our conclusions about the passive nature of the four quiescent galaxies.

\subsection{Constraint on the gas mass fractions}\label{subsec:result_fgas}

We estimate the upper limits on the molecular gas masses of the four quiescent galaxies and Hyde using the upper limits on the [{\sc C i}] line fluxes and the following equation \citep{papadopoulos04,bothwell17}:

\begin{multline}
    M_{[\textsc{C i}]} {\rm (H_2)} = 1375.8\ \frac{D_L^2}{1+z} \left(\frac{X_{[\textsc{C i}]}}{10^{-5}}\right)^{-1} \left(\frac{A_{10}}{10^{-7} {\rm s^{-1}}}\right)^{-1} \\
     \times Q_{10}^{-1}\ S_{[\textsc{C i}]}\Delta v\ [\rm M_\odot],
\end{multline}

\noindent
where $D_L$ is the luminosity distance in Mpc, 
the [{\sc C i}]/$\rm H_2$ abundance ratio $X_{[\textsc{Ci}]}=3\times10^{-5}$, 
the Einstein A coefficient $A_{10}=7.93\times10^{-8} {\rm s^{-1}}$, 
and the excitation factor $Q_{10}=0.6$ \citep{bothwell17}.

We then calculated the upper limits on the molecular gas mass fractions ($f_{\rm gas}=M_{\rm gas}/(M_{\rm gas} + M_*)$) with the molecular gas mass upper limits estimated above and the stellar masses from \citetalias{Schreiber18_mosfire}.
The obtained $3\sigma$ upper limits on the molecular gas masses and gas mass fractions of the individual targets are summarized in Table~\ref{tab:obssummary}.
All of our targets have gas mass fractions of $<0.2$.
The molecular gas mass and gas mass fraction upper limit ($3\sigma$) of the stacked sample (Section~\ref{subsec:stacking_ci}) are $\rm log(M_{gas}/M_\odot)<9.69$ and $f_{\rm gas} < 0.09$.

We compare the gas mass upper limits from [{\sc C i}] of ZF-COS-19589 and Hyde with the gas masses estimated from $L_{\rm IR}$.
In the IR SED library of \citetalias{schreiber18_dustsed}, each SED template has a ratio of $L_{\rm IR}$ and dust mass ($M_{\rm dust}$). 
With this ratio, we can convert the obtained $L_{\rm IR}$ to $M_{\rm dust}$, and then convert $M_{\rm dust}$ to $M_{\rm gas}$ using the gas-to-dust mass ratio as done by \citet{magdis21}. 
Here we use the gas-to-dust mass ratio at solar metallicity \citep{magdis21}. 
Note that $M_{\rm gas}$ obtained with this method is the atomic$+$molecular hydrogen gas mass. 

The gas mass of ZF-COS-19589 inferred from its $L_{\rm IR}$ is $\rm log(M_{gas}/M_\odot)=$ 9.23 and 10.12 with $T_{\rm dust}=$ 40~K and 20~K, respectively.
These values are consistent with the gas mass upper limit from [{\sc C i}] (Table~\ref{tab:obssummary}).
In the case of Hyde, the gas mass inferred from $L_{\rm IR}$ ($T_{\rm dust}=31\pm3$~K) is $\rm log(M_{gas}/M_\odot)=10.45\pm0.3$, which is $0.25$~dex larger than the gas mass upper limit from [{\sc C i}] (Table~\ref{tab:obssummary}).
This could be partly due to a non-negligible contribution from the dust emission heated by old stars on $L_{\rm IR}$ of Hyde \citepalias{schreiber21}. 
Given the uncertainty, however, the gas mass inferred from $L_{\rm IR}$ is consistent with the gas mass upper limit from [{\sc C i}] for Hyde.

\begin{figure}[tbp]
    \centering\includegraphics[width=1.0\columnwidth]{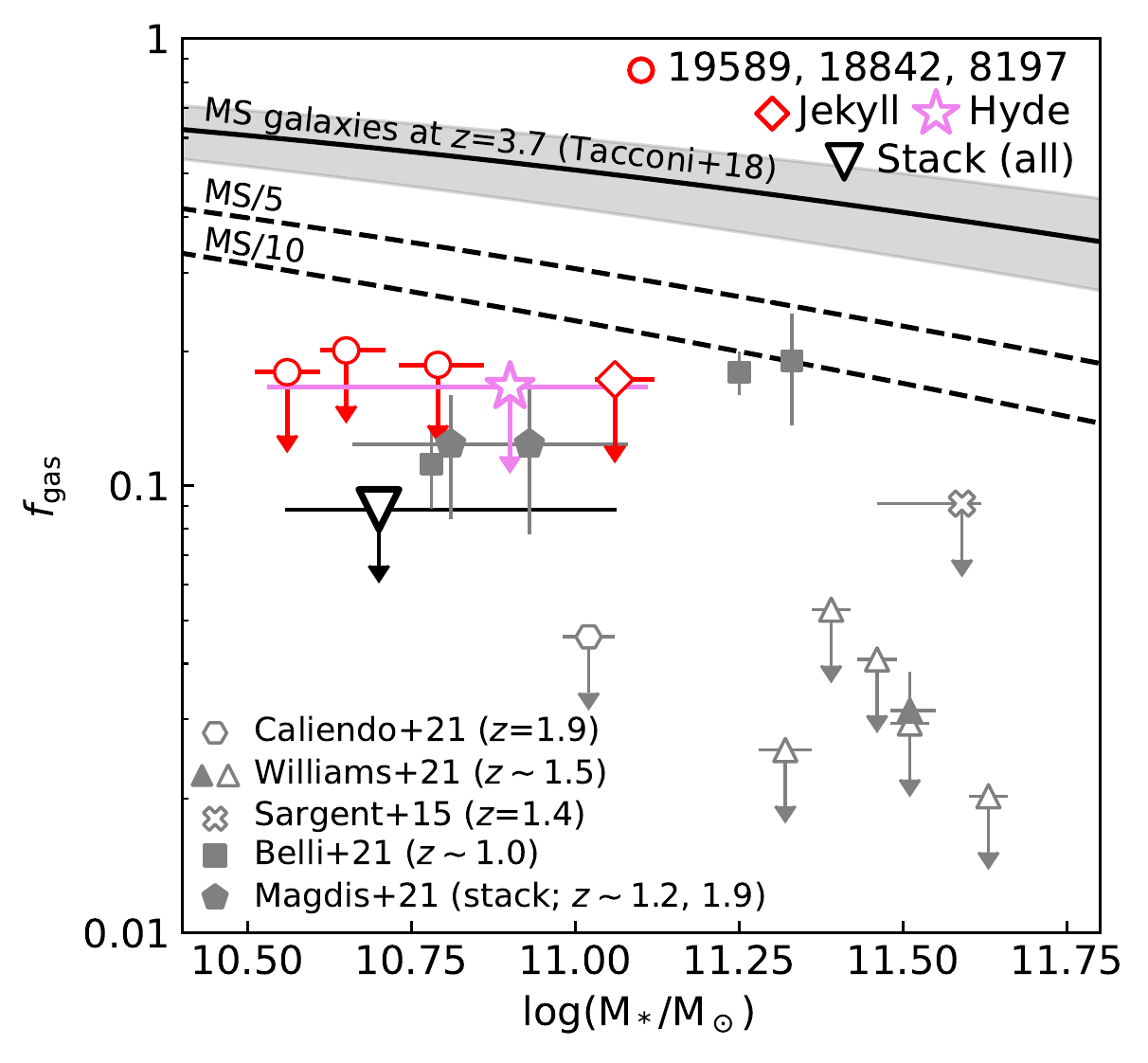}
    \caption{Relation between stellar mass and molecular gas mass fraction upper limit of our targets. 
    Here we show the stacking result for all five galaxies.
    We also show quiescent galaxies at $z=$ 1--3 from the literature \citep{sargent15,belli21,caliendo21,magdis21,williams21} for comparison. 
    The solid line shows the scaling relation between $M_*$ and $f_{\rm gas}$ for galaxies on the star-forming main sequence at $z=3.7$ from \citet{tacconi18}.
    The two dashed lines represent the scaling relations when galaxies are located $5\times$ and $10\times$ below the main sequence.  
    Our individual galaxies and the stacking result at $z=$ 3.5--4.0 have more than three times lower molecular gas mass fractions as compared to the scaling relation for the main sequence galaxies at $z=3.7$. }
    \label{fig:fgas}
\end{figure}

Figure~\ref{fig:fgas}, we show the scaling relation between $M_*$ and $f_{\rm gas}$ for galaxies on the star-forming main sequence at $z=3.7$ from \citet{tacconi18}. 
Comparing with this scaling relation, we find that the gas mass fractions of the five massive galaxies in this work are more than three times lower than main sequence galaxies at the same epoch with similar stellar masses (at $3\sigma$). 
The stacking result is located $6\times$ below the scaling relation for main sequence galaxies. 
These results suggest that massive galaxies at $z=$ 3.5--4.0 consume or expel most of the gas before or during the quenching phase.

\subsection{Constraint on the gas depletion timescales}\label{subsec:tdep}

We estimate the upper limits on the gas depletion timescale ($M_{\rm gas}$/SFR) of ZF-COS-19589 and Hyde with confirmed SFRs (Section~\ref{subsec:SFRIR}). 
The gas depletion timescale of Hyde is estimated to be $<0.3$~Gyr. 
This is shorter than the typical depletion timescale of main sequence galaxies ($\sim$0.4--0.6~Gyr) expected from the scaling relation of \citet{tacconi18}. 
Combining with the upper limit on the gas mass fraction, Hyde appears to be losing almost all its gas content on a short timescale. 
This galaxy might be nearing the end of a starburst phase after which it will become quiescent.
As for ZF-COS-19589, the gas depletion timescale upper limit ($3\sigma$) becomes $<0.4$~Gyr and $<2.4$~Gyr when assuming $T_{\rm dust}=$ 40~K and 20~K, respectively. 
In order to give a further constraint on the depletion timescale of ZF-COS-19589, we would need to constrain its $T_{\rm dust}$. 

\citetalias{Schreiber18_mosfire} characterized the star-formation histories of the quiescent galaxies at $z=$ 3--4 with the parameter $z_{\rm quench}$, the redshift when SFR drops down to 10\% of the SFR in the main formation phase ($\rm <SFR>_{main}$), and $z_{\rm form}$, the redshift when half of the total stellar mass was formed. 
Here, the main formation phase is determined as the contiguous time period surrounding the time of peak SFR where 68~\% of the integrated SFR took place. 
$\rm <SFR>_{main}$ is the mean SFR during this formation phase \citepalias{Schreiber18_mosfire,schreiber18_jekyllhyde}.
The interval between $z_{\rm form}$ and $z_{\rm quench}$ can be used as a proxy of the quenching timescale of galaxies. 
Our four quiescent galaxies dropped down to 10~\% of $\rm <SFR>_{main}$ in 0.15--0.51~Gyr since their formation. 
The star-formation histories from \citetalias{Schreiber18_mosfire} and the low gas mass fractions obtained in this study, therefore, suggest that the molecular gas exhaustion in our galaxies happened on a short timescale of the order of 100~Myr. 
This is consistent with the gas depletion timescale of Hyde. 
\added{These results imply that massive galaxies at $z>3.5$ would have a gas depletion timescale with the order of 100~Myr when they start quenching and keep such a short depletion timescale at least during the quenching phase.}

\added{
It is not clear how long galaxies can keep a short gas depletion timescale after quenching. 
\citet{williams21} suggested the possibility that galaxy quenching happens with a drop in gas mass fraction due to high star-formation efficiency or feedback, which is followed by a period of low gas mass fraction and long gas depletion timescale. 
Indeed, \citet{martig13} showed that the star-formation efficiency of elliptical galaxies with $f_{\rm gas}\sim$ 1.3~\% and 4.3\% is $\sim5$ and $1.2$ times lower than that of spiral galaxies with the same gas mass fraction, respectively, in their simulations. 
It is suggested that the dynamical stabilization of a gas disc by a central bulge becomes effective once the gas mass fraction of a galaxy becomes sufficiently low \citep{martig09,martig13}. 
If our quiescent galaxies at $z=$ 3.5--4.0 are already gas-poor such as $f_{\rm gas}\sim$ 1~\%, they may have a gas depletion timescale with the order of 1~Gyr  \citep{gobat18,magdis21} even though the depletion timescale was shorter when their quenching happened. 
}

\subsection{Comparison with quiescent galaxies at $z<3.5$}\label{subsec:comparison_lowzqgs}

We show quiescent galaxies at $z\sim$ 1--3 from previous studies  \citep{sargent15,belli21,caliendo21,magdis21,williams21} in Figure~\ref{fig:fgas} for comparison. 
Our results on the gas properties of quiescent galaxies at $z>3.5$ are consistent with \citet{sargent15,caliendo21,williams21} and also \citet{whitaker21}, showing that the low star-formation activity of massive quiescent galaxies is due to lack of fuel rather than reduced star-formation efficiency.

\begin{figure}[tbp]
\centering\includegraphics[width=1.0\columnwidth]{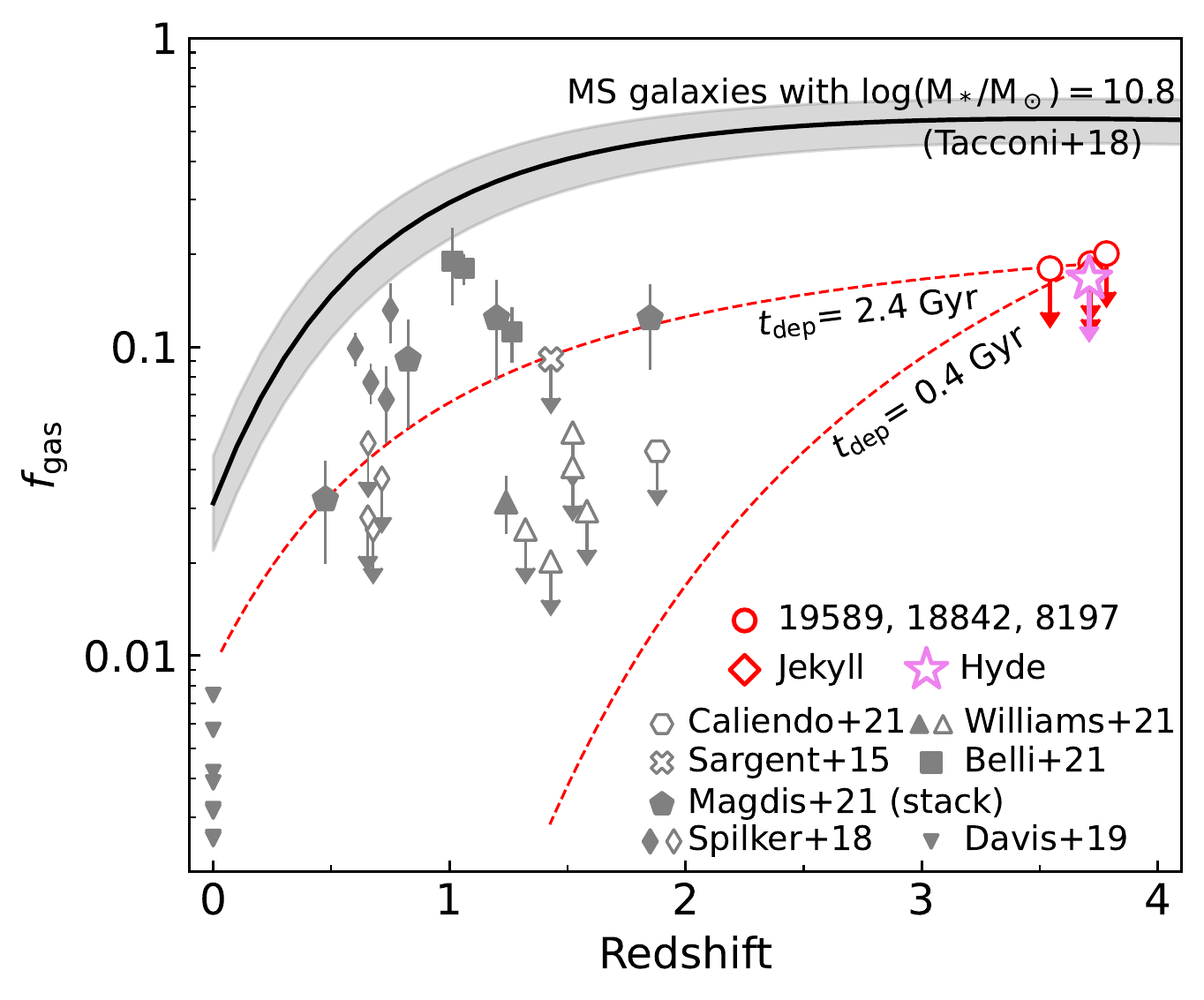}
\caption{Gas mass fraction of quiescent galaxies obtained by this study and previous studies \citep{sargent15,spilker18,davis19,belli21,caliendo21,magdis21,williams21} as a function of redshift.
The thick solid line represents the evolution of gas mass fraction for star-forming galaxies on the main sequence with $\rm log(M_*/M_\odot)=10.8$ from \citet{tacconi18}. 
The shaded area corresponds to the scatter of the main sequence relation (0.3~dex). 
The red dashed line shows the evolution of gas mass fraction of the quiescent galaxies at $z\sim3.7$ when assuming the closed box model and the gas depletion timescale of 0.4~Gyr \added{and 2.4~Gyr}.
}
\label{fig:fgas_evolution}
\end{figure}

Figure~\ref{fig:fgas_evolution} shows the redshift evolution of the gas mass fraction of quiescent galaxies from $z=0$ to $z=4$ 
by combining this work with previous studies \citep{sargent15,spilker18,davis19,belli21,caliendo21,magdis21,williams21}. 
In addition to the quiescent galaxies at $z=$ 1--3 already shown in Figure~\ref{fig:fgas}, we show quiescent galaxies at $z\sim0.8$ \citep{spilker18} and $z\sim0$ \citep{davis19}. 
We also show the evolution of the gas mass fraction of star-forming galaxies on the main sequence with $\rm log(M_*/M_\odot)=10.8$ from \citet{tacconi18}. 

In order to investigate a possible evolutionary path of quiescent galaxies at $z\sim3.7$, we calculate the evolution of the molecular gas mass with a closed-box model, i.e., no inflow and outflow, \added{assuming a constant gas depletion timescale} 
\citep{gobat18, spilker18,williams21}. 
We consider a galaxy with $\rm log(M_*/M_\odot)=10.79$ and $\rm log(M_{gas}/M_\odot)=10.15$ at $z=3.715$ as ZF-COS-19589.  
In the closed-box model, a galaxy is considered to consume the remaining gas only by star-formation according to a constant gas depletion timescale without further gas accretion and removal.   
We assume that $\sim$40\% of the mass of stars formed is returned to ISM (for the Chabrier IMF; \citealt{madau14}).
\added{In Figure~\ref{fig:fgas_evolution}, we show the two model tracks assuming two different gas depletion timescales, namely, $t_{\rm dep}=$ 0.4~Gyr and 2.4~Gyr. 
These values come from the $3\sigma$ upper limits of ZF-COS-19589 with $T_{\rm dust}=40$~K and $20$~K  (Section~\ref{subsec:tdep}). 
Note that because the gas mass fractions of our galaxies at $z=$ 3.5--4.0 are all upper limits, the model track in Figure~\ref{fig:fgas_evolution} should be regarded as an upper limit.}

\added{
The evolutionary track with $t_{\rm dep}=0.4$~Gyr} in Figure~\ref{fig:fgas_evolution} appears to be consistent with the gas mass fraction upper limits of quiescent galaxies at $z\sim$ 1.5--2 from \citet{sargent15,caliendo21,williams21}. 
When we just focus on the gas mass fraction values, the quiescent galaxies at $z\sim3.7$ can evolve into a population of quiescent galaxies with little gas content at $z\sim$ 1.5--2  
\added{if they keep a short gas depletion timescale and experience} no further gas accretion. 
\added{The model track with $t_{\rm dep}=2.4$~Gyr shows a flatter evolution and appears to be consistent with the results from \citet{magdis21}. 
However, quiescent galaxies at $z\sim$ 0.5--1.0 with $f_{\rm gas}\sim0.1$ are more gas-rich than the model track with $t_{\rm dep}=2.4$~Gyr. 
The observed gas mass fractions of quiescent galaxies at $z<3$ show a large scatter at a given redshift, and we cannot explain all the data points at $z<3$ with this simple model when assuming a single gas depletion timescale since $z\sim3.7$. 
This may suggest that quiescent galaxies at $z=$ 3.5--4.0 have various gas depletion timescales although there would be the contributions from galaxies quenched at later epoch, i.e., $z<3.5$ \citep{magdis21}.}

\added{
Note that here we just focus on the gas mass fraction values across cosmic time and do not consider the stellar mass evolution. 
Indeed,}
some of the quiescent galaxies at $z=$ 1--2 in the literature are systematically more massive ($\sim0.5$~dex) than the quiescent galaxies at $z\sim3.7$ (Figure~\ref{fig:fgas}).
It is unlikely that the stellar mass of a quiescent galaxy is increased by more than 0.5~dex only with residual star-formation. 
A fair comparison between quiescent galaxies between at $z\sim$ 3.7 and $z\sim$ 1.5 might be difficult with the current samples. 
It would be necessary to increase the sample size of quiescent galaxies at $z>3$ (and at lower redshifts as well), 
\added{and to give a stronger constraint on their gas properties}
in order to further discuss the evolution of the gas contents in quiescent galaxies across cosmic time.

\subsection{Comparison with Jekyll analogs in a semi-analytical model}\label{subsec:comparison_model}

There are some attempts to identify massive quiescent galaxies at $z>3.5$ in semi-analytical models \citep[e.g.,][]{qin17_20115,rong17}. 
Here we focus on \citet{qin17_20115}, using the {\sc meraxes} semi-analytical model, because this model predicts a similar number density of quiescent galaxies at $z=$ 3--4 as the observed one within a factor of two \citepalias{Schreiber18_mosfire}. 
\citet{qin17_20115} found three analogues of Jekyll in {\sc meraxes} and investigated the time evolution of the analogs.  
They showed that the Jekyll analogs experienced an intense star-formation event and black hole growth via galaxy mergers at $z\sim$ 5--6.  
After the mergers, gas cooling in ISM is significantly suppressed by energy radiated from the central black hole. 
Eventually, the Jekyll analogs consume the remaining cold gas with a timescale of 100--300~Myr and then quench. 
The stellar masses of the Jekyll analogs at $z\sim3.71$ are $\rm log(M_*/M_\odot)=$ 10.95--11.01. 
The gas mass fractions of the Jekyll analogs in the simulation are $f_{\rm gas}\sim0.05$ at $z\sim3.71$, which is consistent with the observed upper limits on the gas mass fractions of the four quiescent galaxies and Hyde.
The observed gas properties of our quiescent galaxies and Hyde are likely to be qualitatively consistent with the formation history of the Jekyll analogs shown in \citet{qin17_20115}. 
This result may suggest the idea that AGN feedback plays an important role in the quenching of massive galaxies at high redshift.

We note that {\sc meraxes} cannot fully reproduce the observed properties of massive quiescent galaxies at $z=$ 3.5--4.0 (\citealt{qin17_20115}; \citetalias{Schreiber18_mosfire}).
It is shown that the Jekyll analogs in {\sc meraxes} have a longer duration of the star-formation phase and quench at later times than the observed quiescent galaxies.  
This indicates that further improvements on the models would be required.  
On the observational side, a stronger constraint on the gas mass and gas depletion timescale for individual quiescent galaxies would be necessary for more quantitative comparison with theoretical models.

\section{Conclusion}\label{sec:summary}
We showed the results obtained from sub-mm observations with ALMA of four quiescent galaxies at $z$=3.5--4 and one optical-dark galaxy at $z=3.7$ named Hyde \citepalias{schreiber18_jekyllhyde,schreiber21}. 
With Band-7 and Band-3, we investigated the presence of the dust-obscured star-formation and the molecular gas traced by the [{\sc C i}] line.
We find that the four quiescent galaxies, including one detected with dust continuum, are located below the main sequence of star-forming galaxies at $z=3.7$ by a factor of $>4$ (at $3\sigma$). 
We confirm that the quiescent galaxies have weak or little dust-obscured star-formation. 
This study demonstrates that the previous multi-wavelength photometric analyses and NIR spectroscopic follow-up observations successfully identified true quiescent galaxies among the mass-selected galaxies at $z=$3--4 \citep{spitler14,Schreiber18_mosfire}.

None of the five targets, including Hyde, have detectable [{\sc C i}] line emission. 
The upper limit on their gas mass fractions is estimated to be $<0.2$. 
Comparing with the scaling relation for galaxies on the star-forming main sequence at $z=3.7$ \citep{tacconi18}, the five massive galaxies have more than three times lower gas mass fractions than star-forming galaxies with similar stellar masses. 
Hyde has an upper limit on the gas depletion timescale of $<0.3$~Gyr, which is shorter than a typical value of the main sequence galaxies at the same epoch. 
Although we cannot give a constraint on the gas depletion timescales of three out of the four quiescent galaxies, 
the upper limits of their gas mass fractions obtained by this study 
and the star-formation histories inferred from the SED fitting \citepalias{glazebrook17,Schreiber18_mosfire} support a scenario where massive galaxies at $z=$ 3.5--4.0 would quench with an abrupt exhaustion of the molecular gas rather than due to a reduction in star-formation efficiency.

Given the large variation of molecular gas properties of quiescent galaxies reported at $z<3$ \citep[e.g.,][]{belli21,williams21}, deeper observations of the gas contents for larger numbers of quiescent galaxies at $z>3$ will be necessary to fully understand the quenching of massive galaxies in the early Universe.

\begin{acknowledgements}
We would like to thank an anonymous referee for a careful reading and comments that improved the clarity of this paper. 
KG acknowledges support from Australian Research Council Laureate Fellowship FL180100060. 
GK acknowledges support of the Australian Research Council Centre of Excellence for All Sky Astrophysics in 3 Dimensions (ASTRO 3D), through project number CE170100013.
PAO is supported by the Swiss National Science Foundation through the SNSF Professorship grant 190079 ‘Galaxy Build-up at Cosmic Dawn’. 
Kavli IPMU is supported by World Premier International Research Center Initiative (WPI), MEXT, Japan. 
The Cosmic Dawn Center (DAWN) is funded by the Danish National Research Foundation under grant No. 140.
This paper makes use of the following ALMA data: ADS/JAP.ALMA\#2018.1.00216.S, ADS/JAP.ALMA\#2019.1.01329.S. 
ALMA is a partnership of ESO (representing its member states), 
NSF (USA) and NINS (Japan), together with NRC 
(Canada), MOST and ASIAA (Taiwan), and KASI (Republic of Korea), in cooperation with the Republic of Chile. 
The Joint ALMA Observatory is operated by ESO, AUI/NRAO and NAOJ.
Data analyses were in part carried out on the open use data analysis computer system at the Astronomy Data Center, ADC, of the National Astronomical Observatory of Japan.

\end{acknowledgements}





\end{document}